\documentclass{elsart}
\usepackage{amssymb}
\usepackage{graphicx}
\newcommand\mbf[1]{\mathbf{#1}}
\newcommand\ovl[1]{\overline{#1}}
\newcommand\tsf[1]{\textsf{#1}}
\newcommand\eeq{\end{equation}}
\newcommand\beq{\begin{equation}}
\begin{document}

\begin{frontmatter}
\title{Comment on ``Exposed-Key Weakness of $\alpha\eta$''
[Phys. Lett. A 370 (2007) 131]}
\author{Ranjith Nair}\ead{nair@eecs.northwestern.edu}
\author{ and Horace P.~Yuen}
\address{ Center for Photonic Communication and Computing\\
Department of Electrical Engineering and Computer Science\\
Northwestern University, Evanston, IL 60208} \maketitle
\begin{abstract}
We show that the insecurity claim of the $\alpha\eta$
cryptosystem made by C.~Ahn and K.~Birnbaum in Phys. Lett. A 370
(2007) 131-135 under heterodyne attack is based
on invalid extrapolations of Shannon's random cipher analysis and on an invalid statistical independence assumption. We show, both for
standard ciphers and $\alpha\eta$, that
expressions of the kind given by Ahn and Birnbaum can at best
be interpreted as security lower bounds.
\end{abstract}
\end{frontmatter}

In \cite{ab07}, Ahn and Birnbaum claim to establish, by an
approximate analysis, the information-theoretic insecurity of the
$\alpha\eta$ encryption system \cite{yuen06,nair06,yuen07pla} even
for ciphertext-only attacks in which Eve makes heterodyne
measurements followed by classical processing. While
information-theoretic security in the asymptotic limit against such attacks has been
claimed by us to be unlikely in \cite{nair06,nair06qcmc}, the main
purpose of this comment is to show that the arguments of
\cite{ab07} do \emph{not} establish insecurity of either the asymptotic or finite cases. We prove the asymptotic insecurity of $\alpha\eta$
ciphertext-only attacks conjectured by us in \cite{nair06,nair06qcmc}, and comment on its lack of practical significance. We also give some new
lower bounds on the average number of spurious keys of
$\alpha\eta$ and other random ciphers.

In Section 1, we describe the claim of \cite{ab07} in the light of known results and conjectures to explain that, despite its
quantitative appearance, they have not given a precise claim that can in principle be falsified. We also summarize our position regarding
Shannon's random cipher and the claims of \cite{ab07}.  In Section 2, we review the concepts of `unicity distance' and
average number of spurious keys $\ovl{N}_k$ of a cipher and the
available results on them in the standard cryptography literature.
In Section 3, we extend these results to random ciphers like
$\alpha\eta$. In Section 4, we critique the analysis of Ahn and
Birnbaum in detail. We also show that their approximate expressions can be replaced by rigorous lower bounds (rather than approximate
equalities) of similar form in the light of our results of Section 3 and that these bounds cannot be used to argue
insecurity of any cipher. We also show that, for $\alpha\eta$, a
true unicity point is never reached for finite $n$ under
known-plaintext attacks, making it more
information-theoretically secure than standard ciphers at least for those
attacks, contrary to the claim of \cite{ab07}. Some concluding remarks are given in
Section 5.

\section{Background and the Claim of \cite{ab07}}

Some specific security analyses and claims on $\alpha\eta$ have been given in \cite{yuen06,nair06,yuen07pla,nair06qcmc}. In particular,
we have expressed \cite{nair06,nair06qcmc} our belief that $\alpha\eta$ in its original form is not information-theoretically secure
under ciphertext-only and known-plaintext attack for large enough $n$. Let $H(K|\mbf{Y}_n)$ be Eve's key uncertainty given the $n$-length
ciphertext $\mbf{Y}_n$. In other words, we claimed without a proof that, even for ciphertext-only attacks, we would have
\beq \label{asymptoticinsecurity}
\lim_{n \rightarrow \infty} H(K|\mbf{Y}_n) = 0.
\eeq
We sketch a proof of this result for ciphertext-only attacks here. Since statistical and known-plaintext attacks give Eve more
information, (\ref{asymptoticinsecurity}) should be expected to hold for these attacks as well. An LFSR gives a periodic output of period
$2^{|K|} -1$ bits. In consequence, observation of the heterodyne attack output over each such running key period provides Eve with
successive observations of the same key corrupted by independent noise coming from the totally random data. At worst, Eve can make her
optimum estimate of the key in each period and take a majority vote at the end. Intuitively, her probability of success goes to unity as
the number of periods goes to infinity in the same way as the average of many measurements of some quantity with independent noise in
each measurement tends to the true value as the number of measurements goes to infinity.

Note however that \emph{ eq. (\ref{asymptoticinsecurity}) has no practical implication} on the security of $\alpha\eta$ in real use since
it is merely an asymptotic statement; see the discussion in ref. \cite{yuen06}. In particular, the PRNG embedded into $\alpha\eta$ is
\emph{not} to be used longer than its period $2^L$ as in the case of standard ciphers, where $L$ is the register length, thus rendering the above insecurity argument inapplicable. For this reason,
and the fact that the attack just mentioned is hugely inefficient, we have not gone into a detailed proof.
Ahn and Birnbaum in \cite{ab07} implicitly make the same claim (\ref{asymptoticinsecurity}) along with a purported proof based on analogy
with Shannon's random cipher \cite{shannon49}. We stress that the argument just given is completely independent of that of Ahn \&
Birnbaum, who provide no evidence for it beyond the analogy with Shannon's random cipher, regarding which we will outline our position
below.

Against this background regarding (\ref{asymptoticinsecurity}), the only new claim with any quantitative justification in \cite{ab07} is
their approximation
\beq \label{approxkeyequivocation}
H_E(K) \approx L - QU \hspace{5mm}\textrm{for} \hspace{5mm}Q \ll n_0 = L/U,
\eeq
where $H_E(K)$ is Eve's equivocation on the key, $L$ the seed key length, $Q$ the length of the data sequence, and $U$ their upper bound
on Eve's information per data bit. By analogy with the Shannon random cipher, the authors then claim ``Eve can determine $K$ with high
probability when ..''
 \beq \label{largeQ}
Q(U+1) \gg L + H_E(K) \eeq
and thus ``the $\alpha\eta$ protocol is worse than the simple additive stream cipher''.

We find the extrapolation from (2) to (3) completely unwarranted and that (3) itself has no more clear meaning than (1). Firstly, since
there is no commonly agreed meaning of the symbols ``$\approx$'' and ``$\ll$'', these statements are not well-defined. As they stand, they
\emph{cannot be falsified}, the possibility of the latter being the hallmark of a meaningful scientific statement. More significantly for
our purpose, there is no reason why (2) is a good approximation in any sense while there is reason to think that it is not, as we will show in detail in this Letter.

Ahn and Birnbaum's argument supporting (\ref{approxkeyequivocation}), which is a heuristic one and not a proof, is again based on a Shannon random cipher analogy which they suggest would be applicable if the PRNG used satisfies a certain pairwise independence condition between any two running key values.

On the one hand, we discuss in detail in Section 4  why their pairwise independence condition is unlikely to lead to an approximate
satisfaction of (\ref{approxkeyequivocation}) in whatever sense and degree they mean, which they have not specified.

On the other hand, we argue that an appeal by analogy to Shannon's random cipher ensemble cannot establish insecurity of any concrete
cipher. Since Shannon's argument uses a large ensemble of ciphers, the average behavior of this ensemble cannot be expected to resemble
that of a given concrete cipher. In consequence, we would not consider any security or insecurity claim that is essentially based on a
Shannon random cipher analogy to be reliable. While we have not examined the issue in detail, we believe that the agreement to this model
observed by Shannon for his `unicity distance' for some concrete ciphers for encrypting English is likely the result of the very special
statistics of English (or any other natural language) that make any cipher encrypting English text quite weak.

However, we attempted to extract the possible meaning and identify the possible validity of the claim (\ref{approxkeyequivocation})
above. It turns out that such a possible rendering of (\ref{approxkeyequivocation}), \emph{similar in form but not in content}, has been
given before \cite{hellman77,bb88} for nonrandom ciphers without the necessity of appealing to Shannon's random cipher assumptions. In
Section 3, we extend the results of Hellman \cite{hellman77} and Beauchemin and Brassard \cite{bb88} (HBB) to random ciphers (in the
sense of Section 2 (see also \cite{nair06}), \emph{not} that of Shannon) like $\alpha\eta$ and analyze it along this rigorous rendering,
different though it is from Ahn and Birnbaum's claim.

\section{Average number of Spurious Keys $\ovl{N}_k$ and `Unicity distance'}

The general form of a random cipher consists of an
encryption map $E_k(\cdot)$ applied by the sender Alice to
a \emph{plaintext} $n$-sequence $\mbf{X}^n = X_1 \ldots X_n$ of
symbols each picked from an alphabet $\mathcal{X}$ resulting in a
\emph{ciphertext} $n$-sequence $\mbf{Y}^n = Y_1 \ldots Y_n$:
\begin{equation} \label{randomencryption}
\mbf{Y}^n = E_k(\mbf{X}^n,\mbf{R}^n). \end{equation}

with the ciphertext symbols belonging to a possibly different
alphabet $\mathcal{Y}$. Note that the encryption map is indexed by
the \emph{secret key} selected randomly from a possible set of
values $\mathcal{K}$ and known only to Alice and the receiver Bob and that the ciphertext is not determined by the key
and plaintext alone but rather
requires an additional random variable $\mbf{R}^n$ generated by
Alice for its complete specification.
The key length $|K|$ is typically of the order of a few 100 bits
for standard ciphers like the Advanced Encryption Standard (AES). The
ciphertext may be openly read by the eavesdropper Eve before
reaching Bob, who applies a corresponding decryption map
$D_k(\cdot)$ to recover the plaintext:
\begin{equation} \label{decryption}
\mbf{X}^n = D_k(\mbf{Y}^n).
\end{equation}
Observe that the
decryption map $D_k$ must function without Bob knowing $\mbf{R}^n$. Further details
on random ciphers may be found in \cite{nair06} -- we note here that
a random cipher usually uses a larger ciphertext alphabet so that $\mathcal{Y} \neq
\mathcal{X}$ - the former may even be continuous as it is for
$\alpha\eta$ under heterodyne attack.

Fixing a particular attack on a given cryptosystem, random or
otherwise, means that the
eavesdropper Eve is assumed to know the joint probability
distribution $\textsf{Pr}[\mbf{X}^n \mbf{Y}^n K]$ of the
plaintext, ciphertext, and key, and is in possession of the
corresponding ciphertext random variable $\mbf{Y}^n$. In the case
of $\alpha\eta$, where information is coded into quantum states,
one must additionally specify a quantum measurement whose result
becomes the ciphertext $\mbf{Y}^n$. In the case of $\alpha\eta$
under heterodyne attack, $\mathcal{X}=\{0,1\}$ and $\mathcal{Y}$
is $\mathbb{R}^2$  or $\mathbb{C}$ since the heterodyne
measurement gives two real numbers. Actually, only the argument of
the complex number result is useful to Eve and thus $\mathcal{Y}$
may be taken to be the circle $\mathcal{S}^1$. In this Letter, we
will consider only information-theoretic security (IT security)
and allow unlimited computational power to Eve.

In the cryptography literature, beginning with Shannon
\cite{shannon49}, the `\emph{unicity distance}' has been proposed as a
measure of IT security of a cipher. The concept may precisely be
defined as the smallest length of plaintext for which only one key
value can lead to the observed ciphertext, thus marking the point
where the system is totally broken. Unfortunately, for most data
statistics, there is never a point where the key becomes fixed
with probability one and the choice of a particular unicity point
involves an implicit choice of a probability that is viewed as
`small enough' and must, in our opinion, be specified in any
insecurity claims. In \cite{shannon49}, Shannon estimated the
unicity distance of an ensemble of ciphers satisfying certain
ideal conditions that are in general not satisfied for a given
cipher. Even for Shannon's random cipher (Throughout this Letter,
we will use `\emph{Shannon's random cipher}' to denote the
ensemble of ciphers defined in \cite{shannon49} and `\emph{random
cipher}' to denote any cipher of the form of Eq.
(\ref{randomencryption}). The reader should keep in mind that they are \emph{completely different} concepts), there is no point where the
key is
fixed with probability one. However, the probability that the key
is erroneously determined by Eve at a designated `unicity point'
can be calculated for this case, as has been done by Hellman in
\cite{hellman77} (see Theorem 1 and Corollary 1 therein). This probability
calculation appears extremely difficult to do for any concrete
cipher, random or otherwise.

In view of the generic non-existence of a true unicity point for a
cipher, we prefer to work with a closely related concept defined
by Hellman \cite{hellman77} for this reason and used also by Beauchemin and
Brassard \cite{bb88}. This is the average number of spurious keys
$\ovl{N}_k$ seen by the attacker that we define below following
\cite{bb88}.

Under a given attack, for each ciphertext $\mbf{y}$, we define the
set $K_\mbf{y}$ as:
\begin{equation} \label{K_y}
K_\mbf{y} = \{k \in \mathcal{K}\hspace{3mm} |\hspace{3mm}
\textsf{Pr}[D_k(\mbf{y})]
> 0\}.
\end{equation}
Thus $K_\mbf{y}$ is the set of keys that could give rise to the
observed ciphertext $\mbf{y}$. Since only one of these keys is the
actual one used, the \emph{number of spurious keys} $N_k(\mbf{y})$
is
\begin{equation} \label{N_ky}
N_k(\mbf{y}) = |K_\mbf{y}| -1.
\end{equation}
The \emph{average number of spurious keys} $\ovl{N}_k$ is defined
to be the expectation of $N_k(\mbf{y})$ over $\mbf{Y}$:
\begin{equation} \label{N_k}
\ovl{N}_k = \sum_{\mbf{y}} \textsf{Pr}[\mbf{y}] N_k(\mbf{y})
\end{equation}
Since each $N_k(\mbf{y})$ is non-negative, if $\ovl{N}_k = 0$,
$N_k(\mbf{y})=0$ for all $\mbf{y}$ and the cipher is broken with
probability one.

It is significant to note that a \emph{unicity distance} $n_0(p)$, which gives the
shortest data length $n_0$ from which Eve could determine the key with probability $p$, is a useful operational measure of security that
one may try to determine numerically or bound analytically for various types of attacks. Some special $p$ cases have been obtained
previously for known-plaintext quantum joint attacks \cite{nairthesis} that yield the fundamental security limit. It is also meaningful
to evaluate $n_0(p)$ under heterodyne or other attacks. Indeed, this is being pursued by different groups in Europe, Japan, and the US on
$\alpha\eta$ and similar cryptosystems.

We stress here that we do not consider $\ovl{N}_k$ by itself to be
an operationally meaningful IT security measure, although it may
well provide bounds on such a measure. Among its drawbacks are the
fact that the cardinality alone of each set $K_\mbf{y}$ defined
above gives no feel for the numerical probabilities of its
elements. In addition, the operational meaning of averaging over
$\mbf{y}$ may be questioned. As an example of an operational
security measure closely related to the unicity distance $n_0(p)$, we suggest the following `$\Pi-$ function'
defined, as a function of the data length $n$ for a given attack
on a given cipher as:
\begin{equation} \label{pifunction}
\Pi(n) := \max_{\mbf{y}^n} \max_{k \in K_{\mbf{y}^n}}
\textsf{Pr}[k | \mbf{y}^n].
\end{equation}
Thus, $\Pi(n)$ is Eve's probability on the most likely key
maximized over all possible ciphertext observations of length $n$.
As such, for a chosen $\epsilon$, if it can be shown that $\Pi
\leq \epsilon$ for the data length of operation, the user can be
guaranteed that the system is broken with a probability not
greater than $\epsilon$ no matter what observation Eve gets. In
this Letter, we do not study the $\Pi$-function as a security
measure since the results on $\ovl{N}_k$, both those available and
those proven here, are closer in spirit and content to the claims
of \cite{ab07} and are sufficient to point out the inadequacies in
their arguments.
 $\ovl{N_k}$
can be estimated exactly for the Shannon random cipher and equals
(see \cite{hellman77}):
\begin{equation} \label{srcnk}
\ovl{N_k} = (2^{H(K)}-1)2^{-nD} \doteq 2^{H(K)-nD},
\end{equation}
where $D$ is the per symbol data redundancy in bits, i.e.,
\begin{equation} \label{redundancy}
D := \log_2(|\mathcal{X}|) - \frac{H(\mbf{X}^n)}{n}.
\end{equation}
Note that $\ovl{N}_k$ never becomes exactly zero, so the cipher is
never broken with probability one. However, Shannon took the point
where $\ovl{N}_k =1$ to be the `unicity distance' $n_0$, so that
$n_0 = H(K)/D$ using the approximation in Eq. (\ref{srcnk}).

For the case of an arbitrary endomorphic nonrandom cipher, i.e.,
one for which $\mathcal{X}=\mathcal{Y}$,  the following
result due to Hellman and Beauchemin and Brassard holds (see \cite{bb88}): \\
\newline \textbf{Theorem 1} (HBB result) For any nonrandom cipher with
$\mathcal{X}=\mathcal{Y}$,
\begin{equation} \label{hbb}
\ovl{N_k} \geq 2^{H(K)-nD} - 1,
\end{equation}

Note that, in contrast to Eq.~(\ref{srcnk}), the RHS of
Eq.~(\ref{hbb}) can reach zero. However, since Theorem 1 gives
just a \emph{lower bound} on $\ovl{N}_k$, the vanishing of its RHS
does \emph{not} establish insecurity in any conceivable
definition. The approximate equality of the right-hand sides of
(\ref{srcnk}) and (\ref{hbb}) led Hellman \cite{hellman77} to
state that Shannon ``random ciphers are essentially the worst
possible'' in the sense of having the lowest possible $\ovl{N}_k$.

Under some restricted assumptions that we do not get into here,
Hellman goes further and gives upper bounds on the
\emph{probability} that $N_k \leq m $ for any integer $m$. These
can obviously be translated into lower bounds on the probability
that $N_k > m$. We do not give the expressions here, because the
important point in our context is that, to judge the
\emph{in}security level of a cipher, we would rather be interested
in \emph{upper bounds} on the probability $\textsf{Pr}[N_k > m]$
which are not available in the analyses \cite{hellman77} and
\cite{bb88} or elsewhere.

In sum, the available results on $\ovl{N}_k$ for nonrandom ciphers
are only lower bounds. As such, they cannot \emph{in principle} be
used to establish insecurity of a system, but may conceivably be
used in conjunction with a meaningful security measure to ensure a
certain security level.

\section{Lower Bound on $\ovl{N}_k$ for Random Ciphers}

The HBB result quoted above can be extended to include random
ciphers with arbitrary ciphertext alphabet $\mathcal{Y}$,
including continuous alphabet ciphers such as $\alpha\eta$. We
prove the extended lower bound in this section.

We need the following lemma that is easily established from standard properties of entropy and mutual information:

\textbf{Lemma}: For any cipher with plaintext sequence
$\mathbf{X}^n = X_1 \ldots X_n$, ciphertext sequence $\mbf{Y}^n =
Y_1 \ldots Y_n$, and key $K$, with arbitrary plaintext alphabet
$\mathcal{X}$ and arbitrary ciphertext alphabet $\mathcal{Y}$,
random or non-random,\beq\label{lemma} H(K|\mbf{Y}^n) =  H(\mbf{X}^n) + H(K)
-I(\mbf{X}^n K ; \mbf{Y}^n).\eeq

\textbf{Theorem 2}: For any cipher with plaintext sequence
$\mathbf{X}^n = X_1 \ldots X_n$, ciphertext sequence $\mbf{Y}^n =
Y_1 \ldots Y_n$, and key $K$, with arbitrary plaintext alphabet
$\mathcal{X}$ and arbitrary ciphertext alphabet $\mathcal{Y}$,
random or non-random,
\begin{equation} \label{ny}
\ovl{N}_k \geq 2^{H(K)+n(\log_2|\mathcal{X}|-D)-I(\mbf{X}^n
K;\mbf{Y}^n)} - 1.
\end{equation}
$D$ is defined as before by Eq.~(\ref{redundancy}). Theorem 1 can be recovered from (\ref{ny}) by observing that $
I(\mbf{X}^n K;\mbf{Y}^n) \leq n\log_2|\mathcal{Y}| =
n\log_2{|\mathcal{X}|}$ when $\mathcal{X}=\mathcal{Y}.$
\newline \\
\textbf{Proof}: We proceed as in \cite{bb88}. We have

\begin{eqnarray}
H(K|\mbf{Y}^n) &=& \sum_{\mbf{y}} \textsf{Pr}[\mbf{y}] H(K|\mbf{y})
\leq \sum_{\mbf{y}} \textsf{Pr}[\mbf{y}]\log_2(N_k(\mbf{y}) +1)\\
&\leq& \log_2(\sum_{\mbf{y}} \textsf{Pr}[\mbf{y}] (N_k(\mbf{y}) +1) )
= \log_2(\ovl{N}_k +1).
\end{eqnarray}
The inequality (15) follows from the definition eq.~(\ref{N_ky})
of $N_k(\mbf{y})$ and (16) from the concavity of the log function
\cite{gallager68}. The result follows on substituting for
$H(K|\mbf{Y}^n)$ using Lemma 1 and exponentiating both sides.
$\blacksquare$

The necessary and sufficient conditions for the inequality of Theorem 2 to be
satisfied with equality are:-The keys in the set $K_\mbf{y}$ must be equiprobable for every$\mbf{y}$ and $|K_\mbf{y}|$ must be the same
for all ciphertexts $y$. Intuitively, these constraints would not be satisfied for an arbitrary cipher, so the lower bound cannot be
expected to be tight without a detailed analysis on the given cipher.
We have thus extended the HBB result to random ciphers and observed
that it is still just a lower bound on the average number of
spurious keys and cannot therefore provide a basis for an
insecurity claim.

\section{Application to $\alpha\eta$ and the analysis of Ahn \& Birnbaum}

We assume the description of the $\alpha\eta$ cryptosystem to be
familiar to the reader from \cite{ab07} -- we use essentially the
same notations here. Further details on the system may be found in
\cite{yuen06,nair06,yuen07pla,nair06qcmc}.

In order to estimate the lower bound in Theorem 2, one needs to
estimate $I(\mbf{X}^n K;\mbf{Y}^n)$ for the cipher being studied.
For $\alpha\eta$, it is useful to define a \emph{signal} random
variable $\mbf{S}^n = S_1 \ldots S_n$ as
\begin{equation} \label{signal} \mbf{S}^n = f^{(n)}(\mbf{X}^n,K),
\end{equation}
where $f^{(n)}$ is simply the function of the data $n$-sequence
and the key that outputs the corresponding $n$-sequence of signal
angles on the coherent state circle. $f^{(n)}$ depends on the
particular PRNG (denoted ENC hereafter to conform with usage in earlier papers) used, but its explicit form does not concern us
here. Each $S_i$ is an $M$-ary random variable. Now the ciphertext
$\mbf{Y}^n = Y_1 \ldots Y_n$ is the $n$-sequence of
continuous-variable heterodyne measurements made by Eve, and may
be represented as
\begin{equation}\label{ciphertext}
\mbf{Y}^n=\mbf{S}^n+\mbf{R}^n,
\end{equation}
where $\mbf{R}^n = R_1 \ldots R_n$, and the $\{R_i\}$ are
independent identically distributed random variables having an
approximately Gaussian distribution with zero mean and standard
deviation $\sigma=\frac{M}{2\sqrt{N}}$, $N$ being the mean photon
number of each transmitted coherent state. They represent the
heterodyne measurement noise of each symbol $i$. For this two-step
model of generation of the ciphertext, note that, for each $i$,  $
(X_i K) \rightarrow S_i \rightarrow Y_i$ is a Markov chain, and
hence so is $ Y_i \rightarrow S_i \rightarrow (X_i K)$ and
consequently, $ \mbf{Y}^n \rightarrow \mbf{S}^n \rightarrow
(\mbf{X}^n K).$ Therefore, by the data processing inequality
\cite{gallager68}, we have for all $n$,
\begin{equation} \label{dpi}
I(\mbf{X}^n K; \mbf{Y}^n) \leq I(\mbf{S}^n ; \mbf{Y}^n).
\end{equation}
Let us denote the running key sequence emitted by an arbitrary ENC
seeded with a seed key of length $|K|$ by $\mbf{K'}= K'_1 \ldots
K'_n \ldots$, where each $K'_i$ is of length $\log_2 (M/2)$ bits
-- the length needed to choose a basis on the coherent state
circle. It is clear that the $\{K'_i\}, 1\leq i\leq n$ cannot be
statistically independent beyond a certain $n$ if each segment of the
running key has a uniform marginal distribution (as is the case
for a pseudo-random number generator), since the seed key entropy
is limited to $|K|$ and the running key is a deterministic
function of the seed key. This fact shows that, for an arbitrary
ENC, there exists a running key length $n_\tsf{dep}$ measured in
running-key symbols, beyond which $\{K'_i\}, 1\leq i\leq n$ are
statistically dependent, and that
\begin{equation} \label{ndep}
n_\tsf{dep} \leq |K|/\log_2(M/2)
\end{equation}
for an arbitrary ENC. $n_\tsf{dep}$ is referred to as the
`dependency distance'. When a linear feedback shift register
(LFSR) is used as an ENC, knowing any $|K|$ consecutive bits of
the output running key fixes the seed key and vice versa.
Therefore, for an LFSR, $n_\tsf{dep}=|K|/\log_2(M/2) \equiv
n_\tsf{dep}(LFSR)$. Note also that if the $\{K'_i\}, 1\leq i\leq
n$ are statistically dependent, so are the signal random variables
$\{S_i\}, 1\leq i\leq n$.

\subsection{Ciphertext-only heterodyne attack}

Consider first the case of ciphertext-only heterodyne attack on
$\alpha\eta$, for which $D=0$. Also the plaintext alphabet size
$|\mathcal{X}| = 2$ for $\alpha\eta$. Ahn and Birnbaum calculate
in \cite{ab07}, a quantity $U$, that is, in our notation :
\begin{equation} \label{defU}
U = I(S_i ; Y_i) \hspace{0.5cm} \forall\hspace{0.2cm} i. \end{equation} This
definition makes sense for the LFSR case (it needs a proof in the
general case) because the $\{S_i\}$ do indeed have the same
(in fact, uniform) marginal distributions for each $i$ when the plaintext
is uniformly random. It is also true that $I(\mbf{S}^n ;
\mbf{Y}^n) = nU$ for all $n \leq n_\textsf{dep}$ because, for such
data lengths, the $i$-th signal symbol in the $n$-sequence is
statistically independent of the rest as mentioned above. However,
this information estimate that is linear in $n$ is \emph{not}
valid beyond the dependency distance because the running key has
correlations beyond $n_\textsf{dep}$. The argument in \cite{ab07}
that the ``pseudo-random number generator redistributes Eve's
prior probabilities back to the flat distribution for each new
symbol'' merely makes $U$ of Eq.~(\ref{defU}) well-defined but
does \emph{not} justify the above estimate. It is the \emph{joint}
probability distribution of the $\{S_i\}$ that goes into the
calculation of $I(\mbf{S}^n ; \mbf{Y}^n)$ and not the marginal per
symbol probability distribution. In fact, it follows from Theorem
4.2.1 of \cite{gallager68} that
\begin{equation} \label{infosequenceub}
I(\mbf{S}^n ; \mbf{Y}^n) < nU \hspace{1cm} \forall n >
n_\textsf{dep},
\end{equation}
and the inequality is strict because the $\{S_i\}$ are not
statistically independent. Even if $I(\mbf{X}^n K; \mbf{Y}^n)$ is
taken to be equal to $I(\mbf{S}^n ; \mbf{Y}^n)$ (see (\ref{dpi})), the claim in \cite{ab07} that the former quantity
increases linearly up to $n_0 = |K|/U$ cannot be true. Note that
$U \approx \frac {1}{2} \log_2 N + 1.6 \ll \log_2 M$ in the regime
$\sigma
 = M/(2 \sqrt{N}) \gg 1$ assumed in the calculation of \cite{ab07} and thus $n_0 \gg
n_\textsf{dep}(LFSR)$, and thus we are already well into the
region where (\ref{infosequenceub}) is a strict inequality. This
argument is unchanged for a general ENC by virtue of the
inequality (\ref{ndep}) -- the running-key dependency sets in not
later than it does for the LFSR case.

The key argument which to them would make Shannon's random cipher analysis in the form of Eq.~(2) (or equivalently,
Eq.~(\ref{infosequenceub}) in the form of an equality) applicable is that for two different running key segments $k_s$ and $k_q$ at the
output of the PRNG ``.. values of $K$ which have similar values of $k_q$ will have uncorrelated values of $k_s$ for $s \neq q$.'' From
the Theorem just cited, (\ref{infosequenceub}) is an equality \emph{if and only if} the $\{S_i\}$ from $1$ to $n$ are jointly
statistically independent, a condition that cannot be satisfied for $n > n_\textsf{dep}$. Their condition quoted above seems to be the
strictly weaker one that the key segments need only be pairwise statistically independent. In view of the well-known difference between
pairwise and complete statistical independence of a sequence of random variables, we feel justified in demanding a rigorous proof of  how
Eq. (22) may be ``approximately'' true under their weaker assumption. We have assumed that by the word `uncorrelated' in the quotation
above, Ahn and Birnbaum  mean `statistically independent' although this is not clear from \cite{ab07}. Whatever their meaning of the
term, we \emph{urge} them to prove how and to what degree it leads to an `approximate' satisfaction of the `only if' condition of Gallager's
theorem that renders (\ref{infosequenceub}) an equality.

Therefore, the only conclusion on $I(\mbf{X}^n K; \mbf{Y}^n)$
derivable from the analysis of \cite{ab07} is that
\begin{equation}
I(\mbf{X}^n K; \mbf{Y}^n) \leq nU \hspace{0.5cm} \forall
\hspace{2mm}n.
\end{equation}
Using this in conjunction with Theorem 2 yields the following
lower bound on $\ovl{N}_k$:
\begin{equation}
\ovl{N}_k \geq 2^{H(K)+ n(1-U)} - 1.
\end{equation}
If we choose to find the data length $n_\textsf{`unicity'}$ at
which the  lower bound reads $\ovl{N}_k \geq 0$, we find
\begin{equation} \label{ctan_0}
n_\textsf{'unicity'} = H(K)/(U-1),
\end{equation}
which is claimed in \cite{ab07} to be the `unicity distance' of
$\alpha\eta$, beyond which ``Eve's entropy on the key will
transition from linear decline to asymptotic decay by analogy to
the unicity distance of a classical deterministic cipher...'' It
is also claimed that ``Eve may have enough information to
determine the key with high probability when $n \gg
n_\textsf{'unicity'}.$''

There are several things amiss with such claims. The fact that the
linear decline of Eve's entropy on the key has already ended at
$n_\textsf{dep}$ has been noted. In addition, the analogy
with Shannon's random cipher does not exist. As stressed in
Sections 2 and 3, for concrete ciphers, the only available results
are lower bounds on $\ovl{N}_k$ against which the analysis of
\cite{ab07} is no exception. As a matter of principle, a lower
bound on $\ovl{N}_k$ \emph{cannot} prove \emph{in}security of a
cipher. If Ahn and Birnbaum wish to claim that $\ovl{N}_k$ is
indeed close to zero at $n_\textsf{`unicity'}$, they must show
both the reasons why the bound of Theorem 2 is tight for
$\alpha\eta$ and also why $I(\mbf{X}^n K; \mbf{Y}^n) \doteq nU$ is
a good approximation for $\alpha\eta$ beyond $n=n_\textsf{dep}$.
Also, if $\ovl{N}_k$ is not claimed to be exactly zero (so the key
is not determined with probability one -- it is shown in
Section 4.2 below that $\ovl{N}_k$ for $\alpha\eta$ is never
exactly zero for any finite data length $n$ under known-plaintext
heterodyne attack and consequently also for the weaker
ciphertext-only attack) -- Ahn and Birnbaum need to estimate the
probability with which Eve obtains the key correctly. As per the
discussion of Sections 2 and 3, this probability can be determined
for Shannon's random cipher but has never been done for \emph{any}
standard cipher, let alone $\alpha\eta$. This fact does not make all previous work in cryptography meaningless because the bulk of it is
concerned with complexity-based security under specific attacks and not information-theoretic security which is under consideration here.
Without such a
calculation, a statement like ``Eve may have enough information to
determine the key with high probability when $n \gg
n_\textsf{'unicity'}.$'' is unfalsifiable -- it does not
satisfy the requirement of being a scientific claim over and above (\ref{asymptoticinsecurity}) without quantifying both how high the
probability is and how much greater than $n_\textsf{`unicity'}$
$n$ needs to be.

\subsection{Statistical and Known-Plaintext Attacks}

For general statistical attacks, i.e., those for which
$H(\mbf{X}^n) < n$, Ahn and Birnbaum claim that a simple additive
stream cipher (ASC) is broken ``with high probability'' when
\begin{equation} \label{ascbroken} n-H(\mbf{X}^n) \gg |K|,
\end{equation} and, by comparison, $\alpha\eta$ is broken ``with
high probability'' when
\begin{equation} \label{alphaetabroken}
n(U+1) - H(\mbf{X}^n) \gg |K|.
\end{equation}
These assertions are again justified by analogy to Shannon's
random cipher analysis, and are interpreted as implying that
$\alpha\eta$ is broken at smaller data lengths than the ASC
because of the added factor of $(U+1)$ in equation
(\ref{alphaetabroken}).

As before, since the terms
``high probability'' and ``$\gg$'' have no precise meaning, these claims are \emph{unfalsifiable} until they are quantified. As with
standard ciphers under many statistical attacks, by choosing $n$
large enough, we can drive the probability of finding the key as
close to $1$ as desired. This is the content of eq.~(\ref{asymptoticinsecurity}), but the equations above ostensibly claim \emph{more}
than that. We contend that \emph{what they claim is not well-defined} without quantitative meaning given to ``high probability'' and
``$\gg$''.

Rigorous bounds, different from (\ref{ascbroken}) and (\ref{alphaetabroken}) although similar in form, can be obtained from an
application of our Theorem
2. For the ASC, we have trivially that $I(\mbf{X}^n K; \mbf{Y}^n)
\leq H(\mbf{Y}^n) \leq n$. Substituting this into the RHS of
Theorem 2 gives the lower bound
\begin{equation} \label{asclowerbound}
\ovl{N}_k \geq 2^{H(K)-nD} - 1,
\end{equation}
which is just the HBB result. As we did for ciphertext-only
attacks, setting the lower bound to zero gives the condition
(compare (\ref{ascbroken}) )
\begin{equation}
n -H(\mbf{X}^n) \geq |K|
\end{equation}
that must be satisfied if $\ovl{N}_k =0$. As such, this is simply
a \emph{necessary condition} for $\ovl{N}_k =0$ and does not imply
the latter.

For $\alpha\eta$, using Eq.~(\ref{infosequenceub}) in Theorem 2
and rewriting $D$ in terms of $H(\mbf{X}^n)$ gives the lower bound
\begin{equation} \label{alphaetalowerbound}
\ovl{N}_k \geq 2^{H(K)+H(\mbf{X}^n)-nU} - 1.
\end{equation}
Setting the RHS to zero, gives the necessary condition (compare
(\ref{alphaetabroken}) )
\begin{equation}
nU -H(\mbf{X}^n) \geq |K|
\end{equation}
for $\ovl{N}_k = 0$. It is not a sufficient condition for the
latter, which, as we show below, is never true except at
$n=\infty$ even for known-plaintext attacks. As is the case for
all applications of Theorem 2, there is no proof that $\ovl{N}_k$
approximately equals the RHS of Eq.~(\ref{alphaetalowerbound})
which would be needed to make insecurity claims on its basis.
Again, it is essential to provide estimates of the probability
that the key is found correctly by Eve to prove insecurity.

Indeed, there is no evidence that (25) is valid as an approximate
estimate of `unicity distance'. The numerical result quoted in
\cite{ab07} for the simulation of \cite{donnet06} yields a
`unicity distance' too small by a factor $\sim 300$, which shows
$U\sim 1$ when (25) is used instead of $U\sim 300$. While such
comparison has little meaning when the attack success probability
is not specified, it surely is unreasonable to claim, as in
\cite{ab07}, that such a large discrepancy exists because of the
suboptimal processing used in \cite{donnet06}.

Intuitively, the measurement noise in $\alpha\eta$ would make it
more secure than an additive stream cipher instead of worse as
claimed in \cite{ab07} at least for the case of known-plaintext
attacks where $H(\mbf{X}^n)=0$. In this case, an ASC is broken
with probability $1$ at the nondegeneracy distance $n_d$ defined
in \cite{yuen06}, which is just $n_d = |K|$ for an LFSR. On the
other hand, it is clearly not possible to pin down the seed key at
this $n$ with probability $1$ in the case of $\alpha\eta$. As a
matter of fact, the true unicity point of $\alpha\eta$ using any
ENC, i.e., the point where the key is determined with probability
one, is \emph{infinite} under even known-plaintext attacks. To see
this, note that, irrespective of what ENC is, in the more exact
continuous Gaussian-noise model of the noise $R_i$ used in
\cite{ab07} (as opposed to the wedge approximation used in
\cite{nair06}), there is always a non-zero probability, however
small, that a $\mbf{Y}^n$ that is close to any given $n$-sequence
of signal points on the coherent state circle may arise from any
data sequence $\mbf{X}^n$ and any running key and thus seed key
$K$. Furthermore, a large fraction (in terms of probability) of
such events for Eve occur without giving rise to any detection
error for Bob. In particular, the close approximation to $R_i$
consisting of a continuous probability distribution cut off at
$90^0$ on each side of the signal point $S_i$ would give zero
error for Bob and infinite unicity distance because every allowed
basis $n$-sequence is still possible given the ciphertext, albeit
some are highly unlikely. The above argument shows that the true
unicity point is not reached for any finite $n$. Together with the
fact that $\lim_{n \rightarrow \infty} \ovl{N}_k = 0$ proven in Section 1 we have that the unicity distance is infinite.
This fact that $\ovl{N}_k \neq 0$ for any specified finite
distance underscores the \emph{necessity} of providing probability
estimates to any claims that the system is broken at that
distance. These estimates are not provided in \cite{ab07} and seem
thus far difficult to obtain, although some progress is currently being made by various research groups.

\section{Conclusion}
In conclusion, we have shown, both by arguing the non-existence of an analogy to Shannon's random cipher and by a direct analysis of
their final claim (\ref{approxkeyequivocation}), that the arguments of Ahn and
Birnbaum  do not establish the insecurity of $\alpha\eta$. Rigorously true results similar in form to their expressions are derived as
corollaries of the lower bound on $\ovl{N}_k$ (Theorem 2). It is noted that these results, being lower bounds, cannot in principle
establish insecurity of any system. We also noted the lack of any estimates by Ahn and Birnbaum of the probability that $\alpha\eta$ is
broken at the claimed distance (which they also did not clearly delimit) to be a serious loophole insofar as it makes their claim of
insecurity unfalsifiable.

There are other points mentioned in \cite{ab07} that we disagree
with but cannot get into in any detail here. One concerns the
comparison of $\alpha\eta$ with DSR to an ASC, and another about
the existence of a proven secure concrete BB84 cryptosystem.  While the work
in \cite{ab07} does not throw light on the true security level of
$\alpha\eta$, further efforts in this direction are possible and welcome.

\section{Added Comment}
In their response \cite{ab08} to our Comment, Ahn and Birnbaum abandon the Shannon random cipher analogy argument of their original paper [1] and repeat their other qualitative
argument that PRNG outputs "will mimic those of a true random number generator".
If that is the case, there is no need for all the work on cryptographic
encryption. Given the previously known condition (1) of our Comment, the
problem here is quantitative. For example, for a fixed data bit length $n$ equal
to the seedkey length under known-plaintext attack, a conventional cipher is
broken with probability 1. If the bare $\alpha\eta$ is broken with probability
$\sim 10^{-4}$ irrespective of complexity, it is already a significant improvement.
Since we have given intuitive as well as rigorous arguments on why these
authors' main claim, our (2), is merely a lower bound that can yield no
insecurity conclusion, its validity can only be established by rigorous
quantitative reasoning the authors have \emph{not} provided.

They also give a simulation example for seedkey size = 13 with other fixed
system parameters. It is not spelled out exactly how the reported simulation was
carried out. In particular, we cannot assess whether or not the method of updating the eavesdropper's probabilities in the simulation
uses any of the assumptions the simulation purports to validate. We  do not comment on it not only for this reason but also because their
original paper used only theoretical arguments to support their conclusion. These arguments remain the same, and our theoretical
refutation still stands.  The importance of addressing our theoretical refutation lies in the fact that a single
numerical example cannot validate a general quantitative conclusion. It would
indeed be interesting if a complete numerical study can be carried out for
realistic key sizes to show the dependence of the results on the system
parameters. However, such study appears exceedingly difficult due to the
complexity involved.

The authors do not dispute the unfalsifiability of their claim given by our (2) in the absence of meaning
given to ``$\approx$'' and ``$\ll$''. Regarding their claim given by (3) in our Letter, insofar as it is meant to say something over and
above our (1), the alleged counter-example in their Reply does not make (3) falsifiable because the example does not satisfy (1). The
issue could be easily resolved if these authors just define
their symbols and give the success probability estimate, which they still
have \emph{not} done. There is no analogy between their claim and the examples
they give, their (7) with $\tan x \approx x$ and their (8). The elementary point to be made here is that the approximation error in those cases can be readily estimated rigorously on demand. In contrast,
their main result given by our (2) is merely a lower bound on $H_E(K)$ -- the crux of the matter is that the gap between $H_E(K)$ and the right-hand side is \emph{unknown}.

We cannot go into here the other side issues raised by these authors in their
Reply. We may just mention that the rigorous examination of unicity distance
for given success probability under a given attack is possible and being
pursued by us and other groups.

\section{Acknowledgements} This work was supported by AFOSR under
grant FA9550-06-1-0452.

\bibliographystyle{elsart-num}

\end{document}